# Ion-acoustic Shock and Solitary Waves in Magnetized Plasma with Cairns-Gurevich Distribution Electrons

Rui Huo, Jiulin Du

*Department of Physics, School of Science, Tianjin University, Tianjin 300350, China*

**Abstract** The propagation properties of ion-acoustic solitary and shock waves in the magnetized viscous plasma with nonthermal trapped electrons are investigated. The Cairns-Gurevich distribution as the electron distribution is considered to describe the plasma nonthermality and particle trapping. By adopting the reductive perturbation technique, we derived the nonlinear Schamel-Korteweg–de Vries-Burgers (SKdVB) equation, and then obtained the ion-acoustic shock and solitary wave solutions of the SKdVB equation for different limiting cases. It is found that the impact of nonthermal parameter $\alpha$, external magnetic field $\Omega$, obliqueness $l_z$, wave speed $U_0$, and the ion kinematic viscosity $\eta_0$ can significantly change the characteristics of the shock and solitary waves. These results may be useful for better understanding the propagation of nonlinear structures in space (i.e. Earth's magnetosphere and ionosphere, auroral regions) and laboratory plasma with nonthermal trapped electrons.

**Keywords**：Ion-acoustic waves; Nonthermal plasma; Trapped electrons; Oblique propagation; SKdVB equation

**1. Introduction**

In recent years, nonlinear structures such as solitons, shocks, and double layers have become the focus of theoretical and experimental research in plasma [1-4]. Among various kinds of nonlinear structures, nonlinear ion-acoustic structures in plasma have been investigated by many researchers in different forms [5-8]. It is known that the formation of ion-acoustic solitary waves is due to the balance of nonlinearity and dispersion. However, when the dissipation is presented in a plasma medium, then shock waves are formed. Dissipation is influenced by many factors, such as wave-particle interactions, dust charge fluctuations, anomalous viscosity, and so on [9-11]. To better understand the properties of the solitary waves and shock waves in plasma, we can study the Korteweg–de Vries (KdV) equation and Korteweg–de Vries Burgers (KdVB) equation to describe them by using the reductive perturbation method. In fact, many researchers have made contributions to this area [12-20].

Presence of external magnetic fields can affect characteristics of the nonlinear waves. Some researchers have studied the effects of an ambient external magnetic field on nonlinear shock, and solitary waves in various plasma [21-22]. The viscous effects which depend on the strength of the external magnetic field are no longer neglected when the gyrofrequency of plasma particles becomes comparable with their collisional frequency. The impact of viscous effects on the properties of shock structures has also been investigated by many authors [23-24].

Trapping of particles has attracted great attention in the study of nonlinear structures. The so-called trapped particles are that, some plasma particles are confined to a finite region of phase space where they bounce back and forth [25]. Schamel generalized a KdV-like equation, often called the Schamel equation, for nonlinear waves modified by trapped electrons [26-27]. This provided a general method for studying the nonlinear behavior of plasma waves in the presence of trapped particles. Many theoretical studies have confirmed that the main properties of various nonlinear structures are significantly affected by trapped particles [28-31].

More and more spacecraft observations from different regions of space plasmas have confirmed that the behavior of charged particles deviates significantly from that with a Maxwellian distribution. These charged particles can be better described by various non-Maxwellian distributions, such as the $q$-distribution in nonextensive statistics [32], the kappa distribution [33], the ($r$, $q$) distribution [34], Cairns distribution [35], and so on. These non-Maxwellian distribution functions have been widely applied to study waves and instabilities in space plasmas [36-39]. Recently, many works focused on different kinds nonlinear structures in space plasma in presence of non-Maxwellian trapped particles [40-45]. These non-Maxwellian trapped particles are described by different Gurevich-distributions. Among these distributions, the Cairns-Gurevich distribution can better describe the trapped nonthermal particles in space plasma such as Earth's magnetosphere and ionosphere [46], auroral regions [47], etc. Ouazene and Amour studied the dust-acoustic solitons in a collisionless dusty plasma with Cairns–Gurevich distributed ions [43]. Arab *et al*. investigated the effect of Cairns–Gurevich polarization force on dust-acoustic solitons in dusty plasmas [44]. Sabrina *et al*. discussed the charge fluctuation induced solitary wave anomalous damping in a dusty plasma in the presence of Cairns–Gurevich distributed ions [45]. However, to our knowledge, the effect of Cairns-Gurevich distribution electrons on ion-acoustic shock and solitary waves in a magnetized plasma has never been discussed. Therefore, in this article, we investigate the propagation properties of nonlinear ion-acoustic shock and solitary waves in a magnetized plasma with Cairns-Gurevich distribution electrons.

The paper is organized as follows. In section 2, the expression of nonthermal trapped electrons number density and basic equations are given in the considered plasma system. In section 3, the nonlinear Schamel Korteweg–de Vries Burgers (SKdVB) equation and ion-acoustic shock and solitary waves solutions for different limiting cases are derived. In section 4, the impact of various plasma parameters on shock and solitary waves is analyzed numerically. In section 5, the conclusion is given.

## 2. Governing Equation

We are considering a dissipative magnetized ion-electron plasma consisting of cold fluid ions and trapped nonthermal electrons following Cairns distribution. The external magnetic field is along the z-axis, i.e., $\mathbf{B} = B\hat{z}$. At equilibrium, the quasi-neutrality condition for the plasma reads as $Z_i n_{i0} - n_{e0} = 0$, where $n_{i0}$ and $n_{e0}$ are the unperturbed number density of ions and electrons, respectively, and $Z_i$ is the charge state of the ion component. In this case, the nonlinear dynamics of the ion-acoustic waves can be described by [20]:

$$\frac{\partial n_i}{\partial T} + \nabla \cdot (n_i \mathbf{u}_i) = 0, \quad (1)$$

$$\frac{\partial \mathbf{u}_i}{\partial T} + (\mathbf{u}_i \cdot \nabla)\mathbf{u}_i = -\frac{Z_i e}{m_i}\nabla \phi + \omega_{ci}(\mathbf{u}_i \times \hat{z}) + \eta_i \nabla^2 \mathbf{u}_i \quad (2)$$

$$\nabla^2 \phi = 4\pi e(n_e - Z_i n_i), \quad (3)$$

where the quantities $n_i$ ($n_e$) and $\mathbf{u}_i$ indicate the unnormalized ion (electron) number density and ion fluid velocity, respectively. $\varphi$ is the unnormalized electrostatic wave potential, $m_i$ denotes the ion mass, and $e$ represents the magnitude of electron charge. $\eta_i$ is the ion kinematic viscosity.

Then we introduce a potential energy $e\varphi$ to the three-dimensional Cairns velocity distribution, the distribution of nonthermal electrons can be described as [35]

$$f_e(v_e) = \frac{n_{eo}}{(1+15\alpha)}\left(\frac{1}{2\pi v_{te}^2}\right)^{3/2}\left[1+\alpha\left(\frac{v_e^2}{v_{te}^2}-\frac{2e\phi}{T_e}\right)^2\right]\exp\left(-\frac{v_e^2}{2v_{te}^2}+\frac{e\phi}{T_e}\right), \quad (4)$$

where $\alpha$ is a parameter that represents the quantity of nonthermal electrons, $v_{te}=\sqrt{T_e/m_e}$ is electron thermal velocity and $\varphi$ is the electrostatic potential. We assume the potential field in the plasma varies slowly and the plasma potential well is large enough to trap a part of nonthermal electrons. In this case, employing Schamel's method, the electrons are free electrons when the electrons energy ($E_e = m_e v_e^2/2 - e\phi$) is greater than zero, i.e. $|v_e|>\sqrt{2e\phi/m_e}$, and the electrons are trapped electrons when the electrons energy is smaller than zero, i.e. $|v_e|\leq\sqrt{2e\phi/m_e}$. Therefore, the three-dimensional Cairns velocity distribution that the adiabatic trapped nonthermal electrons follow can be written as [44]

$$f_e(v_e) = \frac{n_{eo}}{(1+15\alpha)}\left(\frac{1}{2\pi v_{te}^2}\right)^{3/2}$$
$$\times\begin{cases}\left[1+\alpha\left(\frac{v_e^2}{v_{te}^2}-\frac{2e\phi}{T_e}\right)^2\right]\exp\left(-\frac{v_e^2}{2v_{te}^2}+\frac{e\phi}{T_e}\right) & \text{for }|v_e|>\sqrt{2e\phi/m_e} \\ 1 & \text{for }|v_e|\leq\sqrt{2e\phi/m_e}\end{cases} \quad (5)$$

Eq. (5) is what is called the Cairns-Gurevich distribution function. The electron number density $n_e$ can be obtained by integrating Eq. (5) over the velocity space, namely

$$n_e = \int_{\infty}^{-\infty}\int_{\infty}^{-\infty}\int_{\infty}^{-\infty} f_e(v_e)dv_x dv_y dv_z, \quad (6)$$

which can be written as in spherical coordinate ($v_e, \theta, \varphi$),

$$n_e = \frac{n_{e0}}{1+15\alpha}\left(\frac{1}{2\pi v_{te}^2}\right)^{3/2}\int_0^{2\pi}d\varphi\int_0^\pi \sin\theta d\theta\left\{\int_0^{\sqrt{2e\phi/m_e}}v_e^2 dv_e\right.$$
$$\left.+\int_{\sqrt{2e\phi/m_e}}^\infty v_e^2\left[1+\alpha\left(\frac{v_e^2}{v_{te}^2}-\frac{2e\phi}{T_e}\right)^2\right]\exp\left(-\frac{v_e^2}{2v_{te}^2}+\frac{e\phi}{T_e}\right)dv_e\right\}. \quad (7)$$

Calculating the integral above and arranging them, electron number density (including nonthermal free electrons and trapped electrons) can be expressed

$$n_e = n_{e0}\left[F_1 \exp\left(\frac{e\phi}{T_e}\right)erfc\left(\sqrt{\frac{e\phi}{T_e}}\right)+F_2\sqrt{\frac{e\phi}{\pi T_e}}\right] \quad (8)$$

with

$$F_1 = 1 - \frac{12\alpha}{1+15\alpha}\frac{e\phi}{T_e}+\frac{4\alpha}{1+15\alpha}\left(\frac{e\phi}{T_e}\right)^2 \quad \text{and} \quad F_2 = 2+\left[\frac{4\alpha}{1+15\alpha}+\frac{4}{3(1+15\alpha)}\right]\frac{e\phi}{T_e}, \quad (9)$$

where $erfc(x)=1-erf(x)=\frac{2}{\sqrt{\pi}}\int_x^\infty e^{-y^2}dy$ is the complementary error function. The first term in Eq. (8) corresponds to free electrons with Cairns distribution, while the trapped electrons are described by the second term. When $\alpha = 0$, we get the three-dimensional Gurevich density expression [26, 27],

$$n_e = n_{e0}\left\{\exp\left(\frac{e\phi}{T_e}\right)\text{erfc}\left(\frac{e\phi}{T_e}\right) + \left[2 + \frac{4}{3}\frac{e\phi}{T_e}\right]\sqrt{\frac{e\phi}{\pi T_e}}\right\}. \tag{10}$$

If we neglect the trapping effects, Eq. (8) will recover the original nonthermal electron density

$$n_e = n_{e0}\left[\left(1 - \frac{12\alpha}{1+15\alpha}\frac{e\phi}{T_e} + \frac{4\alpha}{1+15\alpha}\left(\frac{e\phi}{T_e}\right)^2\right)\exp\left(\frac{e\phi}{T_e}\right)\right]. \tag{11}$$

In the case of the small amplitude limit ($\phi \ll 1$), Eq. (8) can be expanded in power series as

$$n_e = n_{e0}\left[1 + a_1\left(\frac{e\phi}{T_e}\right) + a_2\left(\frac{e\phi}{T_e}\right)^{3/2} + a_3\left(\frac{e\phi}{T_e}\right)^2\right], \tag{12}$$

where

$$a_1 = 1 - \frac{12\alpha}{1+15\alpha}, \quad a_2 = \frac{4(1+3\alpha)}{3\sqrt{\pi}(1+15\alpha)}, \quad a_3 = \frac{1}{2} - \frac{8\alpha}{1+15\alpha}.$$

Then we introduce the following dimensionless physical quantities:

$$n = n_i/n_{i0}, u = u_i/C_i, \hat{\nabla} = \nabla/\lambda_D,$$

$$t = T\omega_{pi}, \Omega = \omega_{ci}/\omega_{pi}, \varphi = \frac{e\phi}{T_e}, \eta = \eta_i/\omega_{pi}\lambda_D^2. \tag{13}$$

where $C_i = (Z_i T_e/m_i)^{1/2}$ is the ion sound speed, $\lambda_D = (T_e/4\pi e^2 Z_i n_{i0})^{1/2}$ is the Debye length, $\omega_{pi} = (4\pi e^2 Z_i^2 n_{i0}/m_i)^{1/2}$ is the ion plasma frequency. The dynamic equations of ion-acoustic waves can be written as

$$\frac{\partial n}{\partial t} + \hat{\nabla}\cdot(n\mathbf{u}) = 0, \tag{14}$$

$$\frac{\partial \mathbf{u}}{\partial t} + (\mathbf{u}\cdot\hat{\nabla})\mathbf{u} = -\hat{\nabla}\varphi + \Omega(\mathbf{u}\times\hat{z}) + \eta\hat{\nabla}^2\mathbf{u}, \tag{15}$$

$$\hat{\nabla}^2\varphi = -n + 1 + a_1\varphi + a_2\varphi^{3/2} + a_3\varphi^2, \tag{16}$$

## 3. Derivation of the nonlinear equations

To describe the properties of small-amplitude ion-acoustic excitations with our plasma model, we study the Schamel Korteweg–de Vries Burgers (SKdVB) equation and its solutions by the reductive perturbation method. We consider the stretched coordinates as [27]

$$\xi = \varepsilon^{1/4}\left(l_x\hat{x} + l_y\hat{y} + l_z\hat{z} - V_p t\right),$$
$$\tau = \varepsilon^{3/4}t, \tag{17}$$

where $\varepsilon$ ($\ll 1$) is a small parameter that measures the system's nonlinearity strength. $l_x$, $l_y$, and $l_z$ are three direction cosines of wave vector $\mathbf{k}$ along the $x$, $y$ and $z$ axes, and thus, $l_x^2 + l_y^2 + l_z^2 = 1$. $V_p$ is the phase speed of ion-acoustic waves normalized by the ion sound speed, i.e. $C_i = (Z_i T_e/m_i)^{1/2}$. The assumptions $\varepsilon^{1/4}$ with the spatial coordinates and $\varepsilon^{3/4}$ with the temporal coordinate in the above transformations are at first introduced by Schamel, which can better smoothly describe the motion of the ion-acoustic waves. Using Eq. (17), the set of equations (14)-(16) becomes that

$$-V_p\varepsilon^{1/4}\frac{\partial n}{\partial \xi} + \varepsilon^{3/4}\frac{\partial n}{\partial \tau} + \varepsilon^{1/4}l_x\frac{\partial(nu_x)}{\partial \xi} + \varepsilon^{1/4}l_y\frac{\partial(nu_y)}{\partial \xi} + \varepsilon^{1/4}l_z\frac{\partial(nu_z)}{\partial \xi} = 0, \tag{18}$$

$$\varepsilon^{3/4}\frac{\partial u_x}{\partial \tau} + \varepsilon^{1/4}\left(-V_p + u_x l_x + u_y l_y + u_z l_z\right)\frac{\partial u_x}{\partial \xi} + \varepsilon^{1/4}l_x\frac{\partial \varphi}{\partial \xi} - \Omega u_y - \varepsilon^{1/2}\eta\frac{\partial^2 u_x}{\partial \xi^2} = 0, \tag{19}$$

$$\varepsilon^{3/4}\frac{\partial u_y}{\partial \tau}+\varepsilon^{1/4}\left(-V_p+u_x l_x+u_y l_y+u_z l_z\right)\frac{\partial u_y}{\partial \xi}+\varepsilon^{1/4}l_y\frac{\partial \varphi}{\partial \xi}+\Omega u_x-\varepsilon^{1/2}\eta\frac{\partial^2 u_y}{\partial \xi^2}=0, \quad (20)$$

$$\varepsilon^{3/4}\frac{\partial u_z}{\partial \tau}+\varepsilon^{1/4}\left(-V_p+u_x l_x+u_y l_y+u_z l_z\right)\frac{\partial u_z}{\partial \xi}+\varepsilon^{1/4}l_z\frac{\partial \varphi}{\partial \xi}-\varepsilon^{1/2}\eta\frac{\partial^2 u_z}{\partial \xi^2}=0, \quad (21)$$

$$\varepsilon^{1/2}\frac{\partial^2 \varphi}{\partial \xi^2}=-n+1+a_1\varphi+a_2\varphi^{3/2}+a_3\varphi^2. \quad (22)$$

Now, by expanding dependent variables $n$, **u** and $\varphi$ in a power series of $\varepsilon$ in near equilibrium as [41]

$$\begin{aligned}
n &= 1+\varepsilon n^{(1)}+\varepsilon^{3/2}n^{(2)}+\cdots, \\
u_x &= \varepsilon^{5/4}u_x^{(1)}+\varepsilon^{3/2}u_x^{(2)}+\cdots, \\
u_y &= \varepsilon^{5/4}u_y^{(1)}+\varepsilon^{3/2}u_y^{(2)}+\cdots, \\
u_z &= \varepsilon u_z^{(1)}+\varepsilon^{3/2}u_z^{(2)}+\cdots, \\
\varphi &= \varepsilon\varphi^{(1)}+\varepsilon^{3/2}\varphi^{(2)}+\cdots, \\
\eta &= \varepsilon^{1/4}\eta_0+\cdots,
\end{aligned} \quad (23)$$

where $u_x^{(1)}$ and $u_y^{(1)}$ are assumed to vary on a slower scale caused by $E\times B$ drift in magnetized plasma. Substituting Eq. (23) into Eqs. (18)-(22), we get the following terms from the lowest order of $\varepsilon$,

$$n^{(1)}=\frac{l_z}{V_p}u_z^{(1)}, \quad u_y^{(1)}=\frac{l_x}{\Omega}\frac{\partial \varphi^{(1)}}{\partial \xi},$$

$$u_x^{(1)}=-\frac{l_y}{\Omega}\frac{\partial \varphi^{(1)}}{\partial \xi}, \quad u_z^{(1)}=\frac{l_z}{V_p}\varphi^{(1)}, \quad n^{(1)}=a_1\varphi^{(1)}, \quad (24)$$

where we have used the boundary conditions, i.e., the first-order perturbations of variables $n$, **u** and $\varphi$ tend to zero when $\xi\rightarrow\pm\infty$. From above, we can get the expression for the phase speed $V_p$, i.e., $V_p=l_z/\sqrt{a_1}$. Considering the next higher order in $\varepsilon$, we obtain the following set of equations

$$-V_p\frac{\partial n^{(2)}}{\partial \xi}+\frac{\partial n^{(1)}}{\partial \tau}+l_x\frac{\partial u_x^{(2)}}{\partial \xi}+l_y\frac{\partial u_y^{(2)}}{\partial \xi}+l_z\frac{\partial u_z^{(2)}}{\partial \xi}=0$$

$$u_y^{(2)}=-\frac{V_p}{\Omega}\frac{\partial u_x^{(1)}}{\partial \xi}, \quad u_x^{(2)}=\frac{V_p}{\Omega}\frac{\partial u_y^{(1)}}{\partial \xi},$$

$$\frac{\partial u_z^{(1)}}{\partial \tau}-V_p\frac{\partial u_z^{(2)}}{\partial \xi}+l_z\frac{\partial \varphi^{(2)}}{\partial \xi}-\eta_0\frac{\partial^2 u_z^{(1)}}{\partial \xi^2}=0,$$

$$n^{(2)}=a_1\varphi^{(2)}+a_2\left(\varphi^{(1)}\right)^{3/2}-\frac{\partial^2 \varphi^{(1)}}{\partial \xi^2}. \quad (25)$$

From equations (24)-(25), we can derive the SKdVB equation for ion-acoustic shock waves in the plasma,

$$\frac{\partial \Phi}{\partial \tau}+A\Phi^{1/2}\frac{\partial \Phi}{\partial \xi}+B\frac{\partial^3 \Phi}{\partial \xi^3}-C\frac{\partial^2 \Phi}{\partial \xi^2}=0, \quad (26)$$

where we have denoted $\Phi=\varphi^{(1)}$. The SKdVB equation is just valid for trapped electron distribution. If we ignore the effect of ion kinematic viscosity and make $\alpha=0$, Eq. (26) recovers to the Maxwellian distribution case [27]. The other coefficients are expressed as,

$$A=-\frac{3}{4}\frac{a_2}{a_1^{3/2}}l_z, \quad B=\frac{l_z}{2a_1^{3/2}}\left(1+\frac{1-l_z^2}{\Omega^2}\right), \quad C=\frac{\eta_0}{2}; \quad (27)$$

$A$, $B$, and $C$ are the nonlinearity, dispersion, and dissipation coefficients, respectively. Figure 1

shows the absolute of the nonlinearity coefficient $A$ increases with the increase of values of $\alpha$ and $l_z$. Figure 2 shows the dispersion coefficient $B$ increases with the increase of values of $\alpha$ but decreases with the increase of values of $l_z$. Combining Eq. (27), it is concluded that the nonlinearity (corresponding to steepness) of ion-acoustic waves is influenced by the nonthermal parameter $\alpha$ and obliqueness $l_z$. And the dispersion (corresponding to broadening) is affected by the nonthermal parameter $\alpha$, external magnetic field $\Omega$ and obliqueness $l_z$. The ion kinematic viscosity $\eta_0$ is only responsible for the dissipation of ion-acoustic waves.

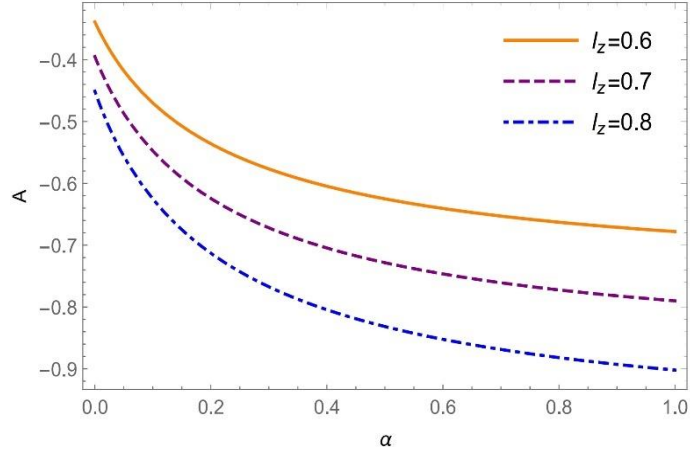

**Figure 1.** The nonlinearity coefficient $A$ as a function of $\alpha$ for different values of $l_z$

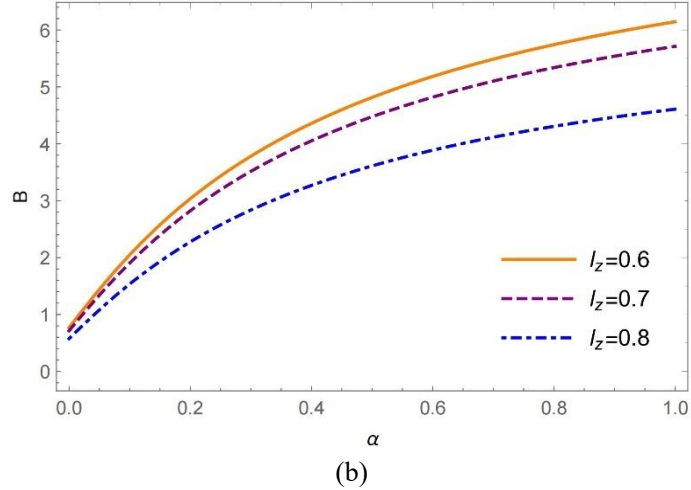

(b)

**Figure 2.** The dispersion coefficient $B$ as a function of $\alpha$ for different values of $l_z$ at $\Omega = 0.5$

**Case I: Presence of ion kinematic viscosity effect**

In the presence of ion kinematic viscosity effect ($C \neq 0$), Eq. (26) leads to the single-shock wave solution [48]. we consider the transformation $\delta = \lambda(\xi - U\tau)$, where $U$ is the velocity of the ion-acoustic shock waves. Using the well-known hyperbolic tangent approach [49] and the appropriate boundary conditions, we obtain the shock wave solution for Eq. (26) in the form (see Appendix A)

$$\Phi(\delta) = \Phi_0 \left(1 + \frac{1}{2}\text{sech}^2(\delta) - \tanh(\delta)\right)^2, \qquad (28)$$

where $\Phi_0 = \left(5C^2/27AB\right)^2$, $U = 28C^2/81B$ and $\theta = \lambda^{-1} = 18B/C$ are the speed and of the width ion-acoustic shock waves, respectively. Using boundary conditions $\Phi = 0$ at $\delta \to +\infty$ and $\Phi = 4\Phi_0$ at $\delta \to -\infty$, we get the amplitude of shock waves in the form

$$\Phi_m = \left(\frac{10C^2}{27AB}\right)^2 \tag{29}$$

**Case II: Absence of ion kinematic viscosity effect**

In the absence of ion kinematic viscosity effect ($\eta_0 = 0$), then $C = 0$, Eq. (26) leads to the solitary wave solution and it becomes:

$$\frac{\partial \Phi}{\partial \tau} + A\Phi^{1/2}\frac{\partial \Phi}{\partial \xi} + B\frac{\partial^3 \Phi}{\partial \xi^3} = 0. \tag{30}$$

To obtain a stationary solution of Eq. (30), we transform the independent variables $\xi$ and $\tau$ to $\chi = \xi - U_0\tau$, where $U_0$ is the speed of the ion-acoustic solitary waves. Using the appropriate boundary conditions, i.e., $\Phi \to 0$, $d\Phi/d\xi \to 0$, $d^2\Phi/d\xi^2 \to 0$ at $\xi \to \pm\infty$, Eq. (30) can be written as

$$-U_0\frac{d\Phi}{d\chi} + A\Phi^{1/2}\frac{d\Phi}{d\chi} + B\frac{d^3\Phi}{d\chi^3} = 0. \tag{31}$$

After some calculation, we can derive the stationary solitary wave solution (see Appendix B), which is given by

$$\Phi = \Phi_m \operatorname{sech}^4\left(\frac{\xi - U_0\tau}{\Delta}\right), \tag{32}$$

where $\Phi_m = (15U_0/8A)^2$ and $\Delta = (16B/U_0)^{1/2}$ are the amplitude and width of ion-acoustic solitary wave.

**4. Numerical analyses**

In this section, we make numerical analyses to show the main characteristics (amplitude and width) of ion-acoustic shock and solitary waves. The effect of nonthermal parameter $\alpha$ and some other plasma parameters (i.e., obliqueness $l_z$, external magnetic field $\Omega$, and the ion kinematic viscosity $\eta_0$) on ion-acoustic shock and solitary wave are discussed.

In Figure 3, we described the variation of ion-acoustic shock waves potential $\Phi$ given in equation (28) as a function of $\delta$ with different parameters. Figure. 3(a) shows the effect of nonthermal parameter $\alpha$ on the ion-acoustic shock waves potential $\Phi$. It is found that the amplitude (width) of shock waves in nonthermal plasma is smaller (wider) than those in Maxwellian plasma. Figure. 3(b) shows how the external magnetic field $\Omega$ affects the shock waves potential $\Phi$. It is observed that the larger parameter $\Omega$ will lead to the larger values of the amplitude of shock waves. Besides, from Eq. (27), we found that with the increase of values of $\Omega$, the dispersion coefficient $B$ decreases so that the width of shock waves is expected to decrease. Figure. 3(c) shows the variation of shock wave potential $\Phi$ with obliqueness $l_z$. It is concluded that as the obliqueness $l_z$ increases, the amplitude of the shock waves decreases and the width increases. Based on the above conclusion, we also can compare the nature of shock waves in magnetized plasma and unmagnetized plasma. When $l_z = 1$, the shock waves in an unmagnetized plasma with trapped nonthermal electrons are recovered. The amplitude (width) of shock waves in an unmagnetized plasma is smaller (wider) than those in a magnetized plasma. Figure. 3(d) shows the impact of viscous coefficient $\eta_0$ on the shock waves. It is clear that with the increase of values of the viscous coefficient $\eta_0$, the amplitude (width) of shock waves increases (decreases). The reason is that the dissipative term dominates over the dispersion term with the increase of values of the viscous coefficient $\eta_0$. From Fig. 3(b) and Fig. 3(d), it is found that the stronger magnetic field contributes to weaken the dispersion and the stronger ion kinematic viscosity contributes to enhance the dissipation, which allows for the formation of larger amplitude and narrower width for shock waves.

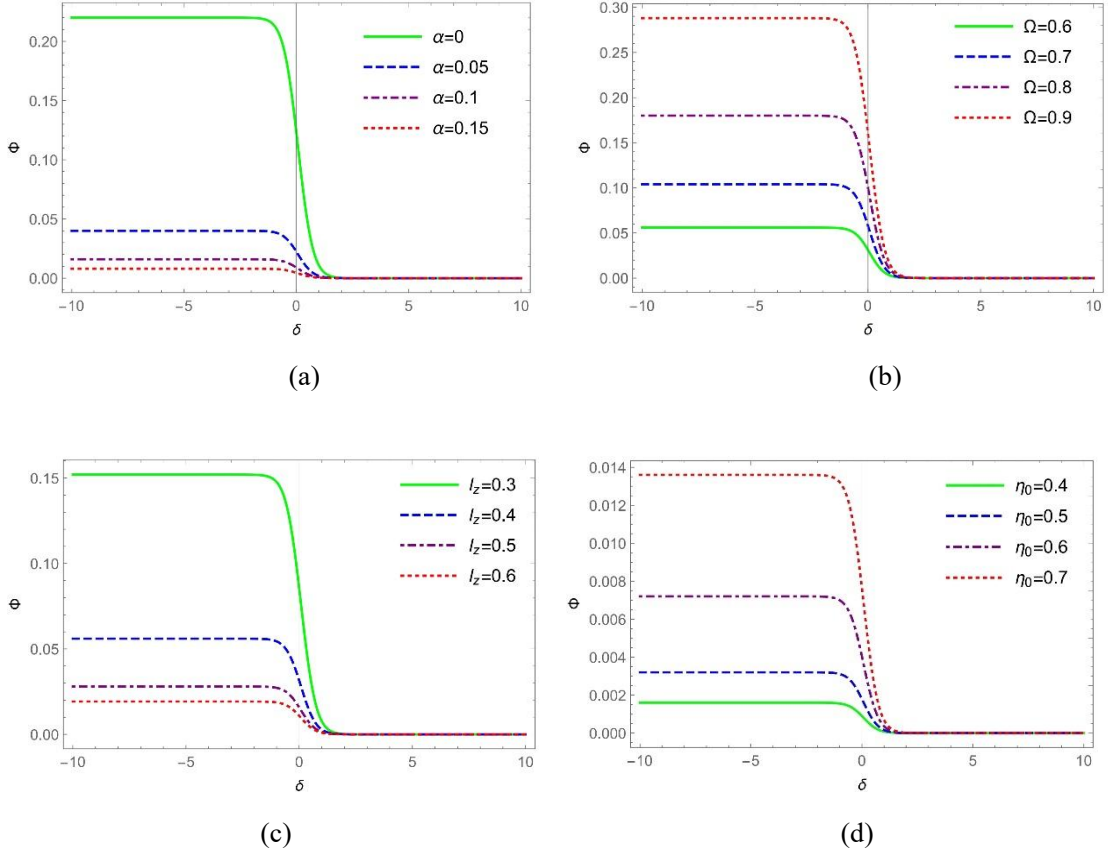

**Figure 3.** Ion-acoustic shock waves potential $\Phi$ as a function of $\delta$ (given in equation (28)) is shown for different values of (a) nonthermal parameter $\alpha$ for $\eta_0 = 0.5$, $\Omega = 0.6$, $l_z = 0.7$. (b) external magnetic field $\Omega$ for $\alpha = 0.1$, $\eta_0 = 0.5$, $l_z = 0.7$. (c) obliqueness $l_z$ for $\alpha = 0.1$, $\Omega = 0.6$, $\eta_0 = 0.5$. (d) viscous coefficient $\eta_0$ for $\alpha = 0.1$, $\Omega = 0.6$, $l_z = 0.7$.

In Figure 4, we described the variation of ion-acoustic solitary waves potential $\Phi$ given in equation (32) as a function of $\chi$ with different parameters. Figure. 4(a) shows the effect of nonthermal parameter $\alpha$ on the ion-acoustic solitary waves potential $\Phi$. It is found that the amplitude of solitary waves decreases but the width of solitary waves increases by increasing the values of nonthermal parameter $\alpha$. This implies that the fewer number of nonthermal electrons, the steeper and the wider the solitary waves. Combining Eq. (27), it is concluded that in the case of solitary structures, nonlinear parameter $A$ is responsible for the amplitude of the waves, while the dispersion parameter $B$ determines the width of the waves. Figure. 4(b) shows the effect of external magnetic field $\Omega$ on the solitary waves. It can be concluded that the magnitude of the external magnetic field will not affect the amplitude of solitary waves. However, the width of solitary waves decreases with the increase of values of the external magnetic field $\Omega$. Fig. 4(c) shows the effect of wave speed $U_0$ on the solitary waves. The increasing value of $U_0$ enhances the amplitude of the solitary waves but makes the width of the waves narrow. Fig. 4(d) shows the impact of obliqueness $l_z$ on solitary waves. It is obvious that with the increase of the values of obliqueness $l_z$, the amplitude of solitary waves decreases but the width of the waves increases. We also can find that the amplitude (width) of solitary waves in an unmagnetized plasma is smaller (wider) than those in a magnetized plasma.

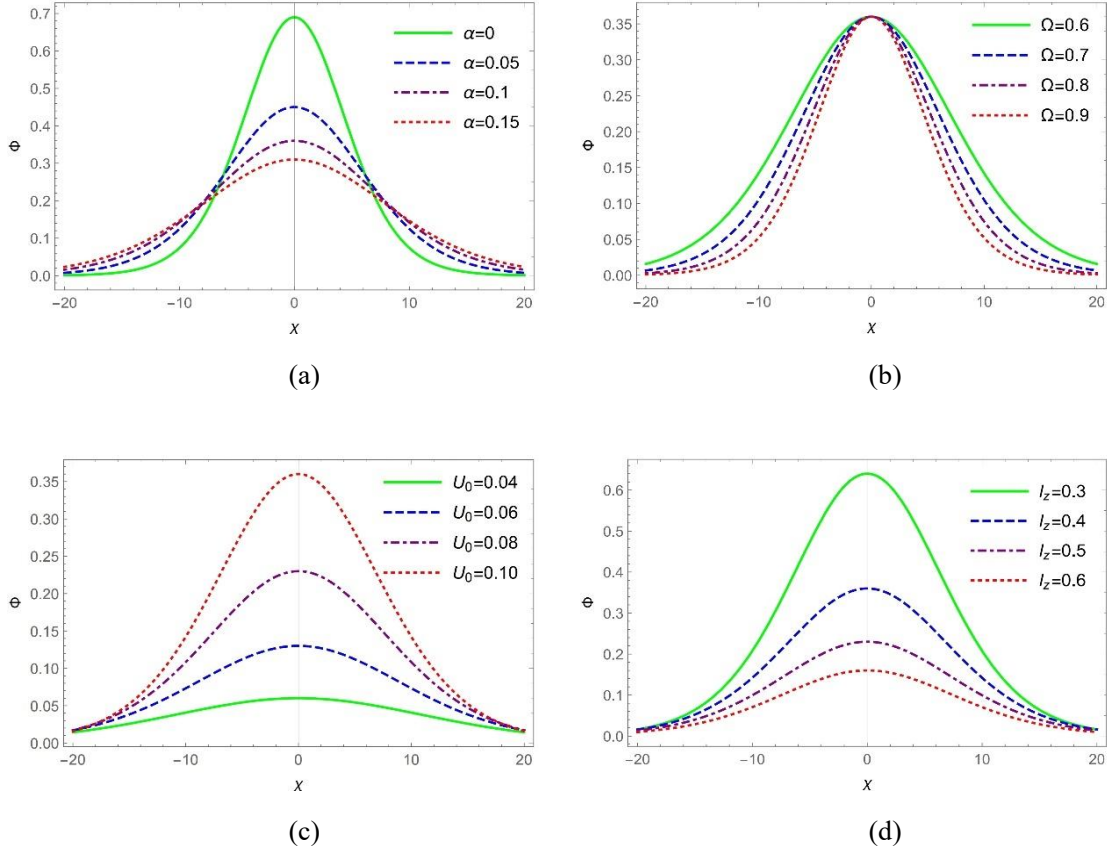

**Figure 4.** Ion-acoustic solitary waves potential $\Phi$ as a function of $\chi$ (given in equation (32)) is shown for different values of (a) nonthermal parameter $\alpha$ for $\Omega = 0.6$, $U_0 = 0.2$, $l_z = 0.4$. (b) external magnetic field $\Omega$ for $\alpha = 0.1$, $U_0 = 0.2$, $l_z = 0.7$. (c) wave speed $U_0$ for $\alpha = 0.1$, $\Omega = 0.6$, $l_z = 0.4$ (d) obliqueness $l_z$ for $\alpha = 0.1$, $\Omega = 0.6$, $U_0 = 0.2$.

## 5. Conclusions

In this paper, the propagation characteristics of ion-acoustic shock and solitary waves have been studied in a magnetized viscous plasma with trapped nonthermal electrons. The expression of trapped electrons number density following Cairns-Gurevich distribution has been reexamined. By adopting the reductive perturbation method, the nonlinear Schamel Korteweg–de Vries Burgers (SKdVB) equation has been derived. Under different limiting cases, the shock and solitary wave solutions of SKdVB have been derived. We found that the impact of nonthermal parameter $\alpha$, external magnetic field $\Omega$, obliqueness $l_z$, wave speed $U_0$, and the ion kinematic viscosity $\eta_0$ may significantly change the characteristics of ion-acoustic shock and solitary waves. The amplitude (width) of ion-acoustic shock and solitary waves is observed to be larger (narrower) in a magnetized Maxwellian plasma than in a magnetized nonthermal plasma. The amplitude (width) of ion-acoustic shock waves is observed to increase (decrease) with the increase of values of external magnetic field $\Omega$. The external magnetic field $\Omega$ has also been observed to modify only the width of ion-acoustic solitary waves. The amplitude (width) for both shock and solitary waves decreases (increases) as the value of obliqueness $l_z$ increases. The higher values of the viscous coefficient $\eta_0$ lead to larger amplitude and narrower width for shock waves.

In summary, the presence of nonthermal [36-39] and trapped particles [28-31] in space and laboratory plasma systems have been reported extensively. Our present work may be useful in understanding the obliquely propagating nonlinear structures (i.e., shock waves, solitary waves) in space plasma (i.e. Earth's magnetosphere and ionosphere [46], auroral regions [47]) and laboratory devices.

**Appendix A**

The KdVB equation (26) can be written as

$$\frac{\partial \Phi}{\partial \tau} + A\Phi^{1/2}\frac{\partial \Phi}{\partial \xi} + B\frac{\partial^3 \Phi}{\partial \xi^3} - C\frac{\partial^2 \Phi}{\partial \xi^2} = 0, \quad (A1)$$

we transform the independent variables $\xi$ and $\tau$ to $\delta = \lambda(\xi - U\tau)$, where $U$ is the velocity of the ion-acoustic waves. Then Eq. (A1) has the form

$$-U\lambda\frac{d\Phi}{d\delta} + A\lambda\Phi^{1/2}\frac{d\Phi}{d\delta} + B\lambda^3\frac{d^3\Phi}{d\delta^3} - C\lambda^2\frac{d^2\Phi}{d\delta^2} = 0. \quad (A2)$$

By setting $\Phi^{1/2}(\delta) = \psi(\delta)$ and $\Phi = \psi^2$, we have

$$-U\frac{d\psi^2}{d\delta} + A\psi\frac{d\psi^2}{d\delta} + B\lambda^2\frac{d^3\psi^2}{d\delta^3} - C\lambda\frac{d^2\psi^2}{d\delta^2} = 0. \quad (A3)$$

By simplifying Eq. (A3) we have

$$-U\psi\frac{d\psi}{d\delta} + A\psi^2\frac{d\psi}{d\delta} + B\lambda^2\left(3\frac{d\psi}{d\delta}\frac{d^2\psi}{d\delta^2} + \psi\frac{d^3\psi}{d\delta^3}\right) - C\lambda\left[\left(\frac{d\psi}{d\delta}\right)^2 + \psi\frac{d^2\psi}{d\delta^2}\right] = 0. \quad (A4)$$

Let $\beta = \tanh\delta$, so that $\frac{d\beta}{d\delta} = 1 - \tanh^2\delta = 1 - \beta^2$, then we have

$$\frac{d}{d\delta} = \frac{d}{d\beta}\frac{d\beta}{d\delta} = 1 - \beta^2\frac{d}{d\beta},$$

$$\frac{d^2}{d\delta^2} = (1-\beta^2)\frac{d}{d\beta}(1-\beta^2)\frac{d}{d\beta},$$

$$\frac{d^2}{d\delta^2} = (1-\beta^2)\frac{d}{d\beta}(1-\beta^2)\frac{d}{d\beta}(1-\beta^2)\frac{d}{d\beta}. \quad (A5)$$

We assume the solution of the KdVB equation has the form $\psi(\delta) = \psi(\beta) = \sum_{n=0}^{N} b_n\beta^n$. The power series breaks off when N = 2 for balancing the highest nonlinearity and dispersive terms [49], thus

$$\psi(\delta) = \psi(\beta) = b_0 + b_1\beta + b_2\beta^2. \quad (A6)$$

Taking Eq. (A5) and (A6) into Eq. (A4), we have

$$
\begin{aligned}
&-U\left(b_{0}+b_{1}\beta+b_{2}\beta^{2}\right)\left(1-\beta^{2}\right)\frac{d}{d\beta}\left(b_{0}+b_{1}\beta+b_{2}\beta^{2}\right)\\
&+A\left(b_{0}+b_{1}\beta+b_{2}\beta^{2}\right)^{2}\left(1-\beta^{2}\right)\frac{d}{d\beta}\left(b_{0}+b_{1}\beta+b_{2}\beta^{2}\right)\\
&+B\lambda^{2}\left[\begin{array}{l}3\left(1-\beta^{2}\right)\frac{d}{d\beta}\left(b_{0}+b_{1}\beta+b_{2}\beta^{2}\right)\times\left(1-\beta^{2}\right)\frac{d}{d\beta}\left(1-\beta^{2}\right)\frac{d}{d\beta}\left(b_{0}+b_{1}\beta+b_{2}\beta^{2}\right)\\ +\left(b_{0}+b_{1}\beta+b_{2}\beta^{2}\right)\left(1-\beta^{2}\right)\frac{d}{d\beta}\left(1-\beta^{2}\right)\frac{d}{d\beta}\left(1-\beta^{2}\right)\frac{d}{d\beta}\left(b_{0}+b_{1}\beta+b_{2}\beta^{2}\right)\end{array}\right]\\
&-C\lambda\left[\begin{array}{l}\left[\left(1-\beta^{2}\right)\frac{d}{d\beta}\left(b_{0}+b_{1}\beta+b_{2}\beta^{2}\right)\right]^{2}\\ +\left(b_{0}+b_{1}\beta+b_{2}\beta^{2}\right)\times\left(1-\beta^{2}\right)\frac{d}{d\beta}\left(1-\beta^{2}\right)\frac{d}{d\beta}\left(b_{0}+b_{1}\beta+b_{2}\beta^{2}\right)\end{array}\right]=0.
\end{aligned}
$$
(A7)

By separating different powers of $\beta$ from Eq. (A7) and solving those equations, one can have different coefficients as

$$b_{0}=\frac{5\left(U+32B\lambda^{2}\right)}{8A},\quad b_{1}=-\frac{10C\lambda}{3A},\quad b_{2}=-\frac{30B\lambda^{2}}{A}.$$
(A8)

Therefore, the solution of Eq. (A2) can be expressed

$$\Phi^{1/2}(\delta)=\frac{5\left(U+32B\lambda^{2}\right)}{8A}-\frac{10C\lambda}{3A}\tanh(\delta)-\frac{30B\lambda^{2}}{A}\tanh^{2}(\delta),$$
(A9)

which can be rewritten as

$$
\begin{aligned}
\Phi^{1/2}(\delta)&=\frac{5}{A}\left(\frac{U}{8}-2B\lambda^{2}-\frac{2C\lambda}{3}\tanh(\delta)+6B\lambda^{2}\left(1-\tanh^{2}(\delta)\right)\right)\\
&=\frac{5}{A}\left(\frac{U}{8}-2B\lambda^{2}-\frac{2C\lambda}{3}\tanh(\delta)+6B\lambda^{2}\operatorname{sech}^{2}(\delta)\right),
\end{aligned}
$$
(A10)

where $\delta=\lambda\left(\xi-U\tau\right)$. By combining the boundary conditions $\Phi=0$ at $\delta\to+\infty$, we get $\lambda=C/18B$ and $U=28C^{2}/81B$. And we can rewrite Eq. (A10) as

$$\Phi^{1/2}(\delta)=\left(\frac{5C^{2}}{27AB}\right)\left(1+\frac{1}{2}\operatorname{sech}^{2}(\delta)-\tanh(\delta)\right).$$
(A11)

Then we can obtain the solution of KdVB Eq. (A2) as

$$\Phi(\delta)=\left(\frac{5C^{2}}{27AB}\right)^{2}\left(1+\frac{1}{2}\operatorname{sech}^{2}(\delta)-\tanh(\delta)\right)^{2},$$
(A12)

where $\delta=\dfrac{C}{18B}\left(\xi-\dfrac{28C^{2}}{81B}\tau\right)$.

**Appendix B**

Eq. (31) can be expressed as

$$-U_{0}\frac{d\Phi}{d\chi}+A\Phi^{1/2}\frac{d\Phi}{d\chi}+B\frac{d^{3}\Phi}{d\chi^{3}}=0.$$
(B1)

Integrating with respect to $\chi$

$$\frac{d^{2}\Phi}{d\chi^{2}}+\frac{2}{3}\frac{A}{B}\Phi^{3/2}-\frac{U_{0}}{B}\Phi=0.$$
(B2)

Multiplying both sides by $(\partial\Phi/\partial\chi)$, we get

$$\frac{d^2\Phi}{d\chi^2}\left(\frac{d\Phi}{d\chi}\right) + \frac{2}{3}\frac{A}{B}\Phi^{3/2}\left(\frac{d\Phi}{d\chi}\right) - \frac{U_0}{B}\Phi\frac{d\Phi}{d\chi} = 0. \tag{B3}$$

Again integrating with respect to $\chi$, we get

$$\frac{1}{2}\left(\frac{d\Phi}{d\chi}\right)^2 + \frac{4A}{15B}\Phi^{5/2} - \frac{U_0}{2B}\Phi^2 = 0, \tag{B4}$$

then

$$\left(\frac{d\Phi}{d\chi}\right)^2 = \frac{U_0}{B}\Phi^2\left(1 - \frac{8A}{15U_0}\Phi^{1/2}\right) \Rightarrow \frac{d\Phi}{d\chi} = \Phi\sqrt{\frac{U_0}{B}\left(1 - \frac{8A}{15U_0}\Phi^{1/2}\right)}$$

$$\Rightarrow \frac{d\Phi}{\Phi} = \sqrt{\frac{U_0}{B}\left(1 - \frac{8A}{15U_0}\Phi^{1/2}\right)}d\chi \Rightarrow \frac{d\Phi}{\Phi\sqrt{\left(1 - \frac{8A}{15U_0}\Phi^{1/2}\right)}} = \sqrt{\frac{U_0}{B}}d\chi. \tag{B5}$$

We make

$$R = \sqrt{\left(1 - \frac{8A}{15U_0}\Phi^{1/2}\right)}, \tag{B6}$$

then we get

$$\Phi = \left(\frac{8A}{15U_0}\right)^2\left(1 - R^2\right)^2, \tag{B7}$$

$$d\Phi = \left(\frac{8A}{15U_0}\right)^2 4\left(1 - R^2\right)(-RdR). \tag{B8}$$

Substituting Eqs. (B7) and (B8) into Eq. (B5), we get

$$-\frac{4dR}{\left(1 - R^2\right)} = \sqrt{\frac{U_0}{B}}d\chi. \tag{B9}$$

Integrating on both sides, we have

$$-\frac{4dR}{\left(1 - R^2\right)} = \sqrt{\frac{U_0}{B}}d\chi$$

$$\Rightarrow -\left[\int\frac{2}{1+R}dR + \int\frac{2}{1-R}dR\right] = \sqrt{\frac{U_0}{B}}d\chi$$

$$\Rightarrow 2\ln(1-R) - 2\ln(1+R) = \sqrt{\frac{U_0}{B}}\chi + K_1 \tag{B10}$$

$$\Rightarrow \ln\frac{(1-R)}{(1+R)} = \frac{1}{2}\sqrt{\frac{U_0}{B}}\chi + \frac{K_1}{2}$$

$$\Rightarrow \frac{(1-R)}{(1+R)} = \exp\left(\frac{1}{2}\sqrt{\frac{U_0}{B}}\chi\right)K_2.$$

From Eq. (B5) we know that at $\chi = 0$ and $\Phi = (15U_0/8A)^2$, which makes $R = 0$. If $R = 0$, we get $K_2 = 1$. Therefore

$$\frac{(1-R)}{(1+R)} = \exp\left(\frac{1}{2}\sqrt{\frac{U_0}{B}}\chi\right). \tag{B11}$$

After some calculations, we get

$$R = \frac{1-\exp\left(\frac{1}{2}\sqrt{\frac{U_0}{B}}\chi\right)}{1+\exp\left(\frac{1}{2}\sqrt{\frac{U_0}{B}}\chi\right)}. \tag{B12}$$

Combining Eqs. (B6) and (B12), we get

$$1-\frac{8A}{15U_0}\Phi^{1/2} = \frac{\left[1-\exp\left(\frac{1}{2}\sqrt{\frac{U_0}{B}}\chi\right)\right]^2}{\left[1+\exp\left(\frac{1}{2}\sqrt{\frac{U_0}{B}}\chi\right)\right]^2}. \tag{B13}$$

$$1-\frac{8A}{15U_0}\Phi^{1/2} = \frac{\left[1-\exp\left(\frac{1}{2}\sqrt{\frac{U_0}{B}}\chi\right)\right]^2}{\left[1+\exp\left(\frac{1}{2}\sqrt{\frac{U_0}{B}}\chi\right)\right]^2} \Rightarrow \frac{8A}{15U_0}\Phi^{1/2} = \frac{\left[1+\exp\left(\frac{1}{2}\sqrt{\frac{U_0}{B}}\chi\right)\right]^2 - \left[1-\exp\left(\frac{1}{2}\sqrt{\frac{U_0}{B}}\chi\right)\right]^2}{\left[1+\exp\left(\frac{1}{2}\sqrt{\frac{U_0}{B}}\chi\right)\right]^2}$$

$$\Rightarrow \Phi^{1/2} = \frac{15U_0}{8A}\frac{4\exp\left(\frac{1}{2}\sqrt{\frac{U_0}{B}}\chi\right)}{\left[1+\exp\left(\frac{1}{2}\sqrt{\frac{U_0}{B}}\chi\right)\right]^2} \Rightarrow \Phi = \left(\frac{15U_0}{8A}\right)^2 \left[\frac{\left[2\exp\left(\frac{1}{4}\sqrt{\frac{U_0}{B}}\chi\right)\right]^2}{\left[1+\exp\left(\frac{1}{2}\sqrt{\frac{U_0}{B}}\chi\right)\right]^2}\right]^2$$

$$\Rightarrow \Phi = \left(\frac{15U_0}{8A}\right)^2 \frac{1}{\left[\frac{\exp\left(-\frac{1}{4}\sqrt{\frac{U_0}{B}}\chi\right)+\exp\left(\frac{1}{4}\sqrt{\frac{U_0}{B}}\chi\right)}{2}\right]^4} \Rightarrow \Phi = \left(\frac{15U_0}{8A}\right)^2 \operatorname{sech}^4\left(\frac{1}{4}\sqrt{\frac{U_0}{B}}\chi\right)$$

$$\Rightarrow \Phi = \left(\frac{15U_0}{8A}\right)^2 \operatorname{sech}^4\left(\frac{\chi}{\sqrt{\frac{16B}{U_0}}}\right) \Rightarrow \Phi = \Phi_m \operatorname{sech}^4\left(\frac{\xi-U_0\tau}{\Delta}\right). \tag{B14}$$